\documentclass[prl,twocolumn,superscriptaddress,english]{revtex4-1}
\usepackage{amsmath,amssymb,amsthm,amsfonts,mathrsfs}
\usepackage{bm}
\usepackage{amstext,bm}
\usepackage{graphicx}
\usepackage{xcolor}
\usepackage{color}
\usepackage{babel}
\usepackage[titletoc]{appendix}
\usepackage{hyperref}
\usepackage{float}
\hypersetup{
        colorlinks = true,
        citecolor  = purple,  
        linkcolor  = blue   
}
\makeatletter

\def \k {\boldsymbol{k}}
\def \H {\mathcal{H}}

\begin{document}

\makeatother
\title{4D spinless topological insulator in a periodic electric circuit}
\author{Rui Yu}
\email[]{yurui@whu.edu.cn}
\affiliation{School of Physics and Technology, Wuhan University, Wuhan 430072, China}
\author{Y. X. Zhao}
\email[]{zhaoyx@nju.edu.cn}
\affiliation{National Laboratory of Solid State Microstructures and
department of Physics, Nanjing University, Nanjing, 210093, China}
\affiliation{Collaborative Innovation Center of Advanced Microstructures, Nanjing University, Nanjing 210093, China}
\author{Andreas P. Schnyder}
\affiliation{Max-Planck-Institute for Solid State Research, D-70569 Stuttgart, Germany}

\begin{abstract}
According to the mathematical classification of topological band structures, there exist a number of fascinating topological states in dimensions larger than three with exotic boundary phenomena and interesting topological responses. While these topological states are not accessible in condensed matter systems, recent works have shown that synthetic systems, such as photonic crystals or electric circuits, can realize higher-dimensional band structures. Here, we argue that the 4D spinless topological insulator, due to its symmetry properties,  is particularly well suited to be implemented in these synthetic systems. We explicitly construct a 2D electric circuit lattice, whose resonance frequency spectrum simulate the 4D spinless topological insulator. We perform detailed numerical calculations of the circuit lattice and show that the resonance frequency spectrum exhibit pairs of 3D Weyl boundary states, a hallmark of the nontrivial topology. These pairs of 3D Weyl states with the same chirality are protected by classical time-reversal symmetry that squares to $+1$, which is inherent in the  proposed circuit lattice. We also discuss how the simulated 4D topological band structure can be observed in experiments.
\end{abstract}
\maketitle

\section*{Introduction}
With the great success of topological band theory in condensed matter physics
\cite{
hasan_RMP_2010,
BOOK_Bernevig_2013,
QXL_RMP_2011,
chiu_Classification_RMP2016,
Weyl_review_2017,
Burkov_Weyl_rev_2018},
recent research has branched out to the study of topological bands in synthetic lattices, such as,
photonic crystals
\cite{
lu_topological_2014,
Photonic_rechtsman_photonic_2013,
LuLing_yan_experimental_2018,
ozawa_TopologicalPhotonics_RMP2019},
ultracold atomic gases
\cite{
ColdAtom_goldman_realistic_2010,
ColdAtom_sun_topological_2012,
ColdAtom_jotzu_HaldaneModel_2014,
ColdAtom_aidelsburger_ChernNumber_Hofstadter_2015,
goldman_TopologicalQuantumMatter_N2016,ZhuSL_Rev},
and electric circuit networks
\cite{
Topological_Circuit_PRX_2015,
Topological_Circuit_PRL_2015,
Topolectrical_Circuits_Weyl_Ronny_CoomP_2018,
luo_topological_2018,
circuit_Weyl_probing_2018,
Topolectrical_Circuits_ssh_PRB_2018,
Thomale_Chern_2019,
Topolectrical_Circuits_Ronny_NP_cornermodes_2018,
hadad_self_inducedTopo_ne_2018,
ezawa_higher_order_PRB2018,
Garcia_LC_PRB2019,
luo_nodal_nodate,li_4Dboundary_2019,
Franz_PRB2019,
Prodan_PRApp_2019,
JieRen_PRB_2019,
liu_topologically_2019,
Ronny_circuitband_PRB2019}.
These synthetic lattices have several advantages compared to their condensed matter counterparts.
One is the ability to precisely control and manipulate the band structure,
another is
the possibility to create lattices in dimensions greater than three.
The celebrated ten-fold classification of topological materials
\cite{
kitaev_PeriodicTableTopological_2009,
Schnyder-PRB-classification,
ryu_Topological_Table_NJP2010,ZYX-PRL-2013,ZYX-PRB-2014}  
predicts a number of interesting higher-dimensional topological states,
including four-dimensional (4D) topological insulators
\cite{Schnyder-PRB-classification,QXL_PRB2008},
4D topological superconductors,
and a 4D generalization of the integer quantum Hall effect \cite{zhang_4D_Science2001}.
These 4D topological states exhibit many interesting phenomena,
e.g.,  quantized nonlinear responses
\cite{zhang_4D_Science2001,froehlich_pedrini_4D_hall_2000,QXL_PRB2008,ryu_moore_PRB2012,lohse_Exploring4DQuantum_Nature2018,price_FourDimensionalQuantumHall_PRL2015},
topological charge pumping,
and in-gap boundary modes with protected level crossings
\cite{kraus_FourDimensionalQuantumHall_PRL2013}.
Unfortunately, these 4D states cannot be realized in condensed matter systems, which are limited to three spatial dimensions.
However, recent technological advances in photonics and cold atomic gases have allowed to synthetically engineer the 4D integer quantum Hall effect,
using, e.g., internal degrees of freedom as additional effective dimensions
\cite{price_FourDimensionalQuantumHall_PRL2015,
price_4D_AI_Topo_2018,
ozawa_SyntheticDimensionsIntegrated_PRA2016,
lohse_Exploring4DQuantum_Nature2018,
petrides_SixdimensionalQuantumHall_nature2018,
zilberberg_PhotonicTopologicalBoundary_Nature2018,
kraus_FourDimensionalQuantumHall_PRL2013}.
These experiments have revealed signatures of charge pumping and topological transport
\cite{lohse_Exploring4DQuantum_Nature2018,zilberberg_PhotonicTopologicalBoundary_Nature2018}.
Apart from these works, there has been no other experimental investigation of the 4D~integer quantum Hall effect, and likewise no other 4D~topological state has yet been realized experimentally.
Among the five 4D topological states of the ten-fold classification
\cite{kitaev_PeriodicTableTopological_2009,
Schnyder-PRB-classification,
ryu_Topological_Table_NJP2010},
the spinless topological insulator, belonging to symmetry class AI, is particularly intriguing.
Its energy bands are characterized by a 4D topological invariant, namely the second Chern number,
which has the distinguishing property of taking on only \emph{even} integer values
\cite{ryu_Topological_Table_NJP2010}.
This invariant leads to topological transport responses in the 4D bulk~\cite{ryu_moore_PRB2012}
and to pairs of Weyl fermions of same chirality on the 3D boundary~\cite{ZYX-PRL-2013,ZYX-PRB-2014}.
Hence, an experimental realization of the 4D spinless topological insulator could
allow to simulate chiral lattice gauge theory of high-energy physics
\cite{Jackiw-Chiral-domain,KAPLAN1992342,BALL19891}.

Besides these interesting properties, the 4D spinless topological insulator has the advantage that
it can be realized easily and in a robust manner in bosonic synthetic or classical systems, such as photonic lattices or periodic electric circuits.
This is because such systems naturally exhibit a time-reversal symmetry that squares
to~$+1$, which is the protecting symmetry of the 4D spinless topological insulator in class AI.
Hence, there is no need to introduce artificial gauge fields or to engineer fine-tuned intra-unit-cell degrees of freedom for the simulation of additional symmetries.
The time-reversal symmetry also guarantees that the first Chern numbers vanish,
such that the topological responses originate purely from the second Chern number.

Motivated by these considerations, we propose in this paper an experimental realization
of the 4D spinless topological insulator in a periodic electric circuit composed of
inductors (L), capacitors (C), and operational amplifiers. By using a mapping between circuit Laplacians and single-particle Hamiltonians, we explicitly construct an circuit lattice,
whose resonance frequency spectrum is identical to a 4D spinless topological insulator in class AI.
We perform detailed numerical simulations of the resonance frequency spectrum for various boundary conditions. For open boundary conditions we observe pairs of 3D Weyl cones that
traverse a gap in the resonance frequency spectrum. Since the LC circuit lattice is non-dissipative, it has a built-in time-reversal symmetry of class AI, which leads to a strong and robust protection of the 3D Weyl boundary states.
Even though the proposed circuit lattice realizes a 4D state, it can readily be implemented on a 2D circuit board or integrated-circuit wafer by projecting the 4D hyperlattice onto the 2D plane.
The crossings of the projected lattice links can be avoided by using a bridge structure for the wiring. The predicted pairs of Weyl modes can be experimentally observed using frequency-dependent measurements.


\section*{4D spinless topological insulator}
We start by discussing a minimal model for the 4D topological insulator in class AI and its boundary Weyl modes. A minimal model can be constructed from a four-band Hamiltonian of the form,
\begin{equation}
\H(\boldsymbol{k})=\sum_{a=0}^{5}f_{a}(\boldsymbol{k})\gamma_{a}.\label{eq:Hk_4D}
\end{equation}
Here $f_a(\k)$ are real functions of the 4D quasi-momentum  $\boldsymbol{k}=(k_{1},k_{2},k_{3},k_{4})$,
{ $\gamma_0 = \bm{1}_{4 \times 4}$,
and $\gamma_i$ (with $i=1,2,\cdots,5$) are five $4\times 4$ gamma matrices,
which satisfy the Clifford algebra $\{\gamma_i,\gamma_j\}=2\delta_{ij}$
and act on the  spinors $\Psi^\dagger=(\psi_a^\dagger,\psi_b^\dagger,\psi_c^\dagger,\psi_d^\dagger)$.
For concreteness we choose the following representation for the gamma matrices: $\gamma_{1,2,3}=\tau_{1,2,3}\otimes\rho_{1}$, $\gamma_4=\tau_0\otimes\rho_2$, and $\gamma_5=\tau_0\otimes\rho_3$, with $\tau_\alpha$ and $\rho_\alpha$ two sets of the Pauli matrices.}
Time-reversal symmetry acts on $\H(\k)$ as $\H^{*}(\k)=\H(-\k)$,
which implies that $f_{0,1,3,5 }$ ($f_{2,4}$) are even (odd) functions of $\k$.
With this condition, one possible choice for $f_i$ that yields a finite second Chern number is:
$f_{0}(\k)=\epsilon-t\cos(k_{2}+k_{3})$,
$f_{1}(\boldsymbol{k})=-t(1+\cos k_{1}+\cos k_{2})$,
$f_{2}(\boldsymbol{k})= t(\sin k_{1}+\sin k_{2})$,
$f_{3}(\boldsymbol{k})=-t(1+\cos k_{3}+\cos k_{4})$,
$f_{4}(\boldsymbol{k})= t(\sin k_{3}+\sin k_{4})$, and
$f_{5}(\boldsymbol{k})=m-t\cos(k_{2}+k_{3})$, similar to a previous model introduced in a general context \cite{price_4D_AI_Topo_2018}. 
Since the term $f_0(\k)$ only affects the global energy at each $\k$,
rather than the topological property as indicated by
$E(\boldsymbol{k})=f_{0}(\mathbf{k})\pm(\sum_{a=1}^{5}f_{a}^{2}(\boldsymbol{k}))^{1/2}$, we choose its form only for the convenience of the realization of the 4D topological electric circuit.

The topology of the gapped 4D class AI system can be characterized by the second Chern number.
For the Dirac model, the second Chern number can be nicely simplified as the winding number of $\hat{\boldsymbol{f}}=\boldsymbol{f}/|\boldsymbol{f}|$ from the 4D Brillouin zone (BZ) to the 4D unit sphere $S^{4}$~\cite{QXL_PRB2008}
	\begin{equation}
	C_{2}=\frac{3}{8\pi^{3}}\int d^{4}k~\epsilon^{\mu\nu\lambda\rho\sigma}\hat{f}_{\mu}\partial_{k_{1}}\hat{f}_{\nu}\partial_{k_{2}}\hat{f}_{\lambda}\partial_{k_{3}}\hat{f}_{\rho}\partial_{k_{4}}\hat{f}_{\sigma},\label{eq:Ch2}
	\end{equation}
where $\epsilon^{\mu\nu\lambda\rho\sigma}$
is the rank-5 Levi-Civita symbol with $\mu,\nu,\lambda,\rho,\sigma=1,2,\cdots,5$,
and repeated indices are summed over.
Straightforward calculation gives that
$C_{2}=-2$ if $-t/2<m<t$, and otherwise $C_{2}=0$  as shown in
Fig.\ref{4D-model} \textbf{a}, for which a detailed derivation can be found in the supplemental materials.

	        \begin{figure}[ht]
	        \includegraphics[width=1.1\columnwidth]{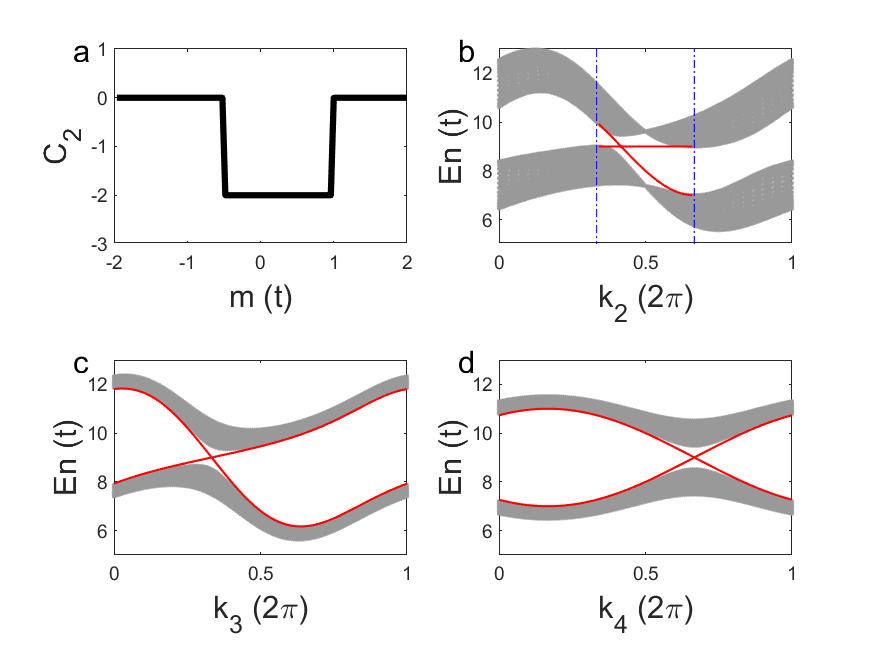}
	        \caption{{\label{fig:1}
	        Topological properties of the 4D model Hamiltonian}.
	        \textbf{a}, The second Chern number $C_{2}$ as a function of $m$ (in unit of $t$). For $-t/2<m<t$, $C_{2}=-2$.
	        \textbf{b}, The band structures for a slab geometry confined in the $\boldsymbol{r}_{1}$ direction, with $m=0$. The $k$-line in the $(k_{2},k_{3},k_{4})$ space is chosen to cross the point $w_{1}=2\pi( 5/12, 1/3,-1/3)$ along the $k_{2}$ direction. The bulk part of the band structure (gray) are obtained by projecting the eigenvalues of the Hamiltonian (\ref{eq:Hk_4D}) with $k_{1}\in [0,2\pi]$. The boundary states (red lines) are the eigenstates of the boundary effective Hamiltonian (\ref{eq:H_surface_3D}) in the range of $k_{2}\in2\pi(1/3,2/3)$, namely between the two vertical lines (blue dashed lines).
	        \textbf{c} and \textbf{d}, The bulk band structures and the boundary Weyl states along the $k_{3}$ and $k_{4}$ directions, respectively. The boundary states appear in the whole range of $k_{3,4}\in[0,2\pi]$. The local band structure around the point $w_2$ is related to that around $w_1$ by time-reversal symmetry.
	        \label{4D-model}
	        }
	        \end{figure}

\begin{figure*}
			\includegraphics[width=1\linewidth]{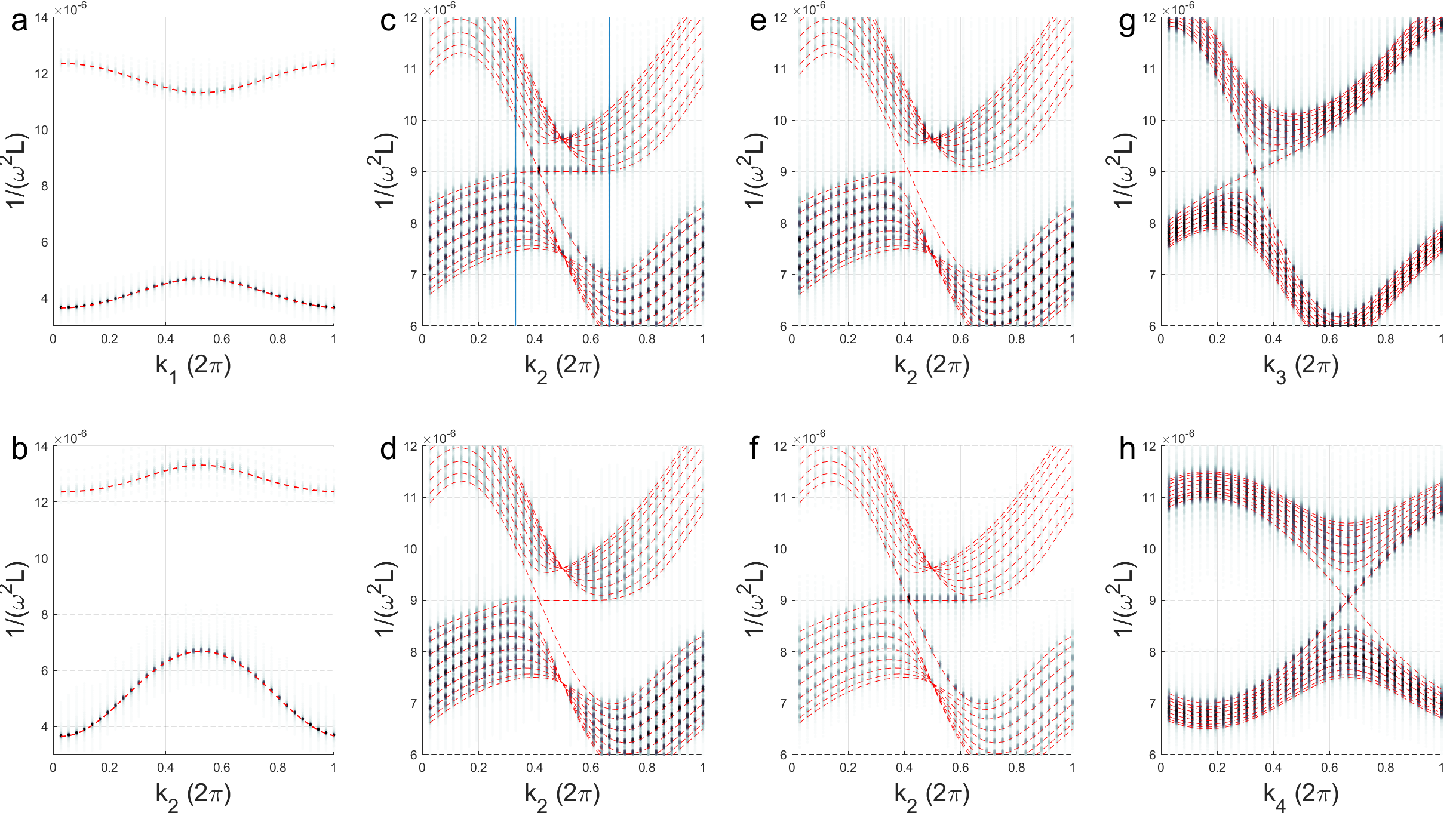}
			\caption{{\label{fig:Simu_result} The band structures for the 4D circuit
			lattice with periodic boundary conditions (PBCs) and open boundary conditions (OBCs).} For each figure, the results obtained from simulation are plotted by gray dots, while in comparison, those from the model Hamiltonian are presented by red dashed lines.
			\textbf{a} (\textbf{b}), The bulk band structure along the $k_{1}$ ($k_{2}$) direction passing through the origin of the BZ for the PBCs.
			\textbf{c}-\textbf{h}, The voltage intensities along various $k$-lines crossing one of the Weyl points, $w_1$, under the OBCs for the $\boldsymbol{r}_{1}$-direction and PBCs for the other directions. In simulation, we assume nine layers for the $\boldsymbol{r}_{1}$-direction.
			\textbf{c}, The voltage intensity of the bottom four layers, contributed from all
			the four types of nodes, along the $k_2$-direction.
			From the intensity distribution, we observe that the in-gap boundary Weyl states exist in the interval from $k_{2}=2\pi/3$
			to $4\pi/3$, as marked by the two vertical (blue) lines. The simulation result agrees with the analytic result from the model Hamiltonian, particularly well in the interval.
			\textbf{d}, The voltage intensity from the 5th layer
			to the top edge layer. The surface Weyl states disappear since the pulse source added on the bottom boundary cannot excite the surface Weyl states on the top boundary, demonstrating the local nature of the surface states.
			\textbf{e} and \textbf{f}, The voltage intensities on the bottom four layers from the
			$a$, $b$ and $c$, $d$ nodes, respectively.
			The intensity appears to be dominated by the $c$ and $d$ components rather than the $a$ and $b$ components, consistent with the analytic result from the model Hamiltonian.
			\textbf{g} and \textbf{h}, The band structures along $k_{3}$ and $k_{4}$ directions, respectively.
			The linear dispersion relations in the vicinity of the $w_1$ point along all directions on the boundary
			demonstrate the $w_{1}$ point is a 3D Weyl point.}
			\end{figure*}

According to the general theory of bulk-boundary correspondence of topological insulators, a nontrivial second Chern number leads to boundary Weyl fermions.
We consider a 3D boundary perpendicular to the $\boldsymbol{r}_{1}$-axis, putting the semi-infinite system in the region with $r_1>0$.
For simplicity we set $m=0$ and the system is in the topologically nontrivial phase with $C_{2}=-2$. For the Dirac model \eqref{eq:Hk_4D} the boundary effective Hamiltonian can be derived analytically as~\cite{mong_edge_2011}
	\begin{equation}
	\H_{s}(\tilde{\k})=f_{0}(\tilde{\k})\sigma_{0}-f_{3}(\tilde{\k})\sigma_{1}+f_{4}(\tilde{\k})\sigma_{2}+f_{5}(\tilde{\k})\sigma_{3}.\label{eq:H_surface_3D}
	\end{equation}
Here $\sigma_\alpha$ acts in the sub-lattices $c$ and $d$,
$f_{0,3,4,5}$ are functions defined in \eqref{eq:Hk_4D}, depending only on $\tilde{\k}\equiv(k_2,k_3,k_4)$, where
$k_{2}\in(2\pi/3,4\pi/3)$ and $k_{3,4}\in[0,2\pi]$.
From the boundary spectrum $E(\tilde{\k})=f_{0}\pm(\sum_{a=3}^{5}f_{a}^{2})^{1/2}$, it is easy to obtain that there are two Weyl points located at
$w_{1,2}=\pm 2\pi( 5/12, 1/3,-1/3)$ in the 3D boundary BZ
as shown in Fig.\ref{fig:1} \textbf{c}-\textbf{d}.
{As the Weyl points are located at generic momenta, they have anisotropic dispersion relations due to the lack of rotational symmetry.
The $\gamma_0$ term in Eq.~\eqref{eq:Hk_4D} leads to the unwanted $\sigma_0$ term in Eq.~\eqref{eq:H_surface_3D}, which tilts the boundary Weyl points. But, we have to make a trade off between the simplicity of the model and the magnitude of the term.
Since the two Weyl points are related by time-reversal symmetry, they have the same chirality, right-handedness as shown in the supplementary materials. More details are given in the supplementary materials. Generically the boundary states decay exponentially towards the bulk, which will be confirmed by our simulation results in Fig.\ref{fig:Simu_result}.}

		\begin{figure*}
		\includegraphics[width=1\linewidth]{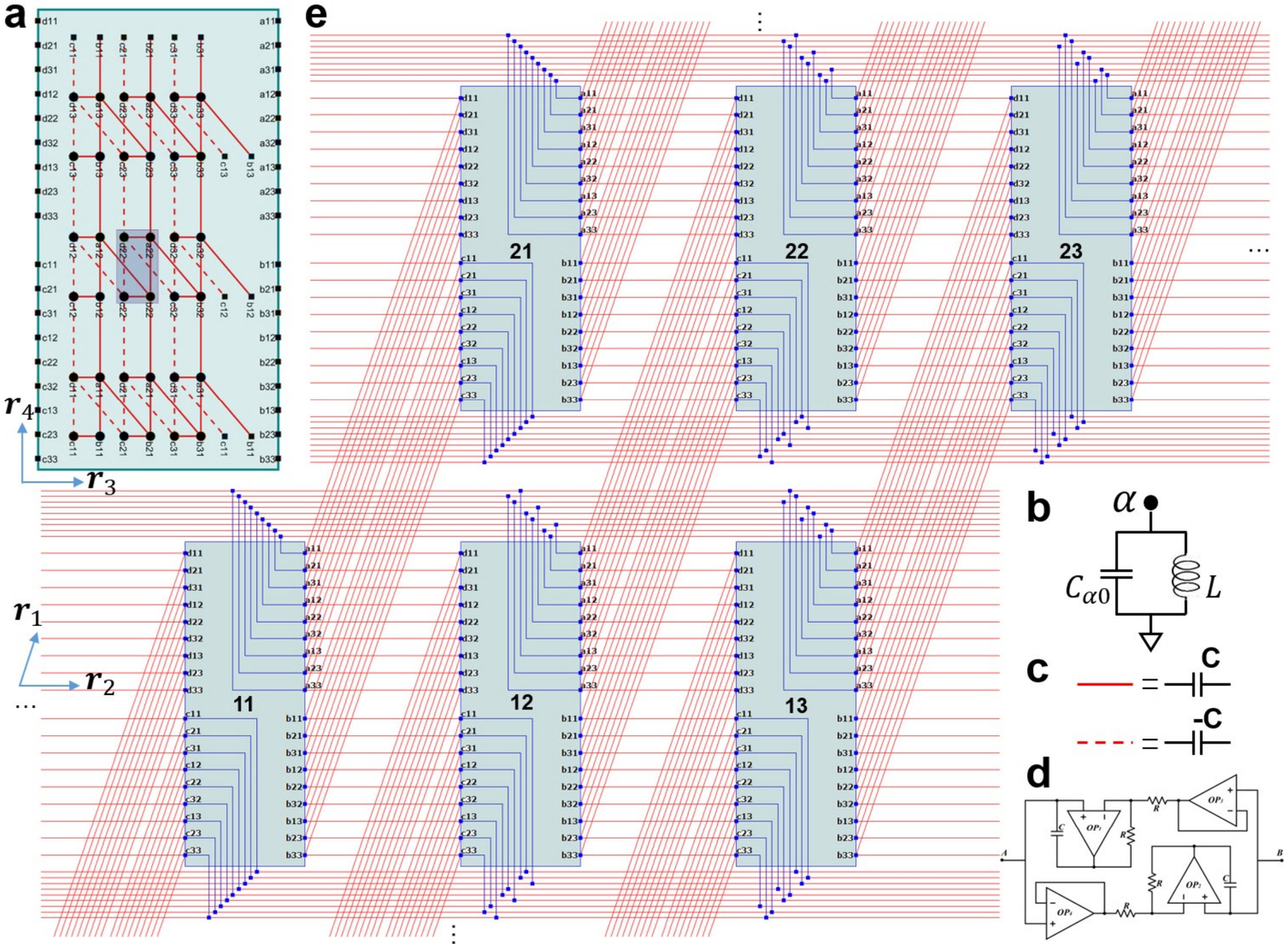}
		\caption{\label{fig_circuit}
		{ The 4D circuit lattice realized on a 2D plane.}
		\textbf{a}, A 2D sub-circuit lattice containing $3\times3$ unit cells in a $\boldsymbol{r}_{3}$-$\boldsymbol{r}_{4}$ plane of the 4D circuit lattice, and \textbf{b} and \textbf{c}, two basic components used in the 2D sub-circuit.  In panel \textbf{a}, the small dark blue rectangle exemplifies a unit cell consisting of four nodes, $a,b,c,d$, denoted by black dots. All nodes on the plane are labeled by $\alpha ij$, where $\alpha=a,b,c,d$, and $i$ ($j$) is the lattice index for the $\boldsymbol{r}_{3}$ ($\boldsymbol{r}_{4}$) direction.
		Each node in the plane is connected to ground through the component illustrated in panel \textbf{b}, which contains a capacitor and an inductor connected in parallel. On the plane, within each unit cell, connections are made between $a$ to $b$, $b$ to $c$, $c$ to $d$, and $d$ to $a$. Node $a$ ($c$) in each unit cell is connected to node $b$ ($d$) in a neighbor cell if the two cells are separated by the vector $(0,0,1,0)$ or $(0,0,0,1)$ [$(0,0,-1,0)$ or $(0,0,0,-1)$].
		The two types of connections are indicated by solid and dashed red lines, respectively. As illustrated in panel \textbf{c}, each solid (dashed) line indicates the component containing a capacitor with capacitance $C$ ($-C$).
		The Born-von Karman periodic boundary conditions are implemented by connecting the nodes on the right (top) edge to the corresponding nodes on the left (bottom) edge.
		In order to facilitate the connections on the ($\boldsymbol{r}_{1}$, $\boldsymbol{r}_{2}$) plane, we connect all nodes to the black squares on the edges with the same indices by wires, which are not explicitly shown to make the figure neat. 	Moreover, black squares with the same label are equipotential.
		\textbf{d}, The two-port sub-circuit as an effective capacitor with capacitance $-C$. A detailed derivation for this result is given in the supplementary materials.
		\textbf{e}, The circuit lattice with $2\times 3$ blocks in the $\boldsymbol{r}_{1}$-$\boldsymbol{r}_{2}$ plane. Here, each block is a copy of the 2D sub-circuit in panel \textbf{a}, with the lattice indices on the $\boldsymbol{r}_{1}$ and $\boldsymbol{r}_{2}$ plane indicated at the center. Each $a$ ($b$) node is connected to a $d$ ($c$) node if they are separated by the vector $(1,0,0,0)$ or $(0,1,0,0)$, and each $a$ ($c$) node is connected to another $a$ ($c$) node if they are separated by $(0,\pm1,\pm1,0)$.
		The blue lines indicate wires, and the solid red lines are again specified in panel \textbf{c}. In addition, wires are connected at a crossing point if it is marked as a blue square. Otherwise, they just go across each other without connection.
	}
		\end{figure*}

\section*{Tight-binding model and circuit lattice}
We now proceed to address the realization of the above 4D topological states
by constructing a realistic electric-circuit in a practical way.
For this purpose, it is more convenient to write the model Hamiltonian~\eqref{eq:Hk_4D} in real space,
that reads
$H=\sum_{\alpha,\beta,i,s}t_{\alpha\beta}(R^s_{\alpha\beta})c_{\alpha}^{+}(i+R^s_{\alpha\beta})c_{\beta}(i)$, where $\alpha, \beta$ label nodes in each unit cell and
$i$ labels the unit cells.
$R_{\alpha\beta}^s$ are hopping vectors, which can be obtained by the inverse Fourier transform of \eqref{eq:Hk_4D}, and are listed as:
$R_{ad}^s= R_{bc}^s=(0,0,0,0),(1,0,0,0),(0,1,0,0)$ and
$R^s_{ab}=-R^s_{cd}=(0,0,0,0),(0,0,1,0),(0,0,0,1)$ with $s=1,2,3$, respectively, and
$R^s_{aa}= R^s_{cc}=(0,1,1,0),(0,-1,-1,0)$ with $s=1,2$, respectively. Here, each number in the parentheses is in the unit of the corresponding lattice constant for the 4D hypercube lattice.
The hopping amplitudes are assumed to be
$t_{ab}=t_{bc}=t_{ad}=t_{aa}=t_{cc}=-t_{cd}=-t$, where $t$ is a real constant so that time-reversal symmetry is preserved.
Exchanging the order of the subscripts, the amplitudes $t_{\alpha\beta}$ are unchanged while the vectors $R_{\alpha\beta}$ are reversed.
The above 4D tight-binding model can be mapped to a 4D circuit lattice as detailed
in the supplementary materials.
As the property of a circuit lattice depends only on the connection relations among its nodes, regardless of the shape of circuit lattice, one can project the 4D circuit lattice onto a 2D plane to obtain an equivalent 2D circuit lattice as shown in
Fig.\ref{fig_circuit}, preserving the property of the circuit.
In more detail, the circuit in Fig.\ref{fig_circuit} is constructed by the following two steps.
First, the sub-circuits in the $\boldsymbol{r}_{3}$-$\boldsymbol{r}_{4}$ planes,
with the Born-von Karman periodic boundary conditions, are constructed as shown in Fig.\ref{fig_circuit} \textbf{a}.
The nodes and lines in Fig.\ref{fig_circuit} \textbf{a} are detailed in Fig.\ref{fig_circuit} \textbf{b}-\textbf{d}.
Then, arrange the sub-circuits on the $\boldsymbol{r}_{1}$ and $\boldsymbol{r}_{2}$ plane, and connect the nodes between sub-circuits with capacitors to realize the connections in the $\boldsymbol{r}_{1}$-$\boldsymbol{r}_{2}$ planes as shown in Fig.\ref{fig_circuit} \textbf{e}.
By the two steps, we have constructed a 2D circuit that is genuinely equivalent to the 4D circuit, since the connections of nodes in the two circuits have a one-to-one correspondence.
According to the Kirchhoff current law, it is easy to check that the current equations for the circuit in Fig.\ref{fig_circuit} \textbf{e} lead to a Hamiltonian with exactly the same form as Eq.~\eqref{eq:Hk_4D}. Now the functions $f_a(\boldsymbol{k})$ have the parameters concretely specified in terms of capacitance values as $t=C$, $m=(C_{a0}-C_{b0}+2C)/2$ and
$\epsilon=(C_{a0}+C_{b0})/2+7C$ as detailed in the supplementary materials.
If the capacitance values satisfy $C_{a0}+2C=C_{b0}$, namely $m=0$,  the circuit is in a topologically nontrivial phase with the second Chern number $C_{2}=-2$.

\section*{Simulation results}
In order to extract the resonance frequency spectrum of the circuit lattice,
we perform the time-domain transient simulations to obtain the voltage $v(t,\boldsymbol{R},\alpha)$ on each node as a function of time.
Here, $\boldsymbol{R}$ is the unit cell label, $\alpha=a,b,c,d$ is the index for the nodes in each unit cell, and $t$ is the time.
Taking periodical boundary conditions in $\boldsymbol{r}_{1,2,3,4}$ directions, respectively, and performing the Fourier transform, the voltage $v(\omega,\boldsymbol{k},\alpha)$ can be obtained in the momentum $\boldsymbol{k}$ and frequency $\omega$ space.
The band-structure-like dispersions are obtained by plotting $|v(\omega,\boldsymbol{k},\alpha)|$.
As introduced in the supplementary materials, the eigenvalue $\varepsilon$ of the tight-binding model corresponds to the resonance frequency $\omega$ of the circuit lattice, with the relation $\varepsilon=1/(\omega^2L)$. Therefore, the vertical axis in Fig.\ref{fig:Simu_result} are plotted as $1/(\omega^{2}L)$, so as to compare with the eigenvalues of the Hamiltonian (\ref{eq:Hk_4D}).
In Fig.\ref{fig:Simu_result}~\textbf{a} and \textbf{b},
it is easy to see that the simulation results (gray points) are in good agreement with the bulk band dispersions (red dashed line) obtained from the model Hamiltonian (\ref{eq:Hk_4D}).

Next we study the surface states by assuming open boundary conditions in the $\boldsymbol{r}_{1}$ direction and periodic boundary conditions in the$\boldsymbol{r}_{2,3,4}$ directions. The technical details are provided in the supplementary materials.
The pulse voltage source is connected to the (1,1,1,1) cell on the bottom edge, and thereby the voltage $v(t,\boldsymbol{R},\alpha)$ are obtained for a slab geometry with nine layers in the  $\boldsymbol{r}_{1}$ direction. We then carry out the Fourier transforms for $\boldsymbol{r}_{2,3,4}$ and $t$, which gives $v(\omega,R_{1},\tilde{\k},\alpha)$, where $R_{1}$ is the lattice index in the $\boldsymbol{r}_{1}$ direction and $\tilde{\k}=(k_2,k_3,k_4)$. The corresponding boundary band structures for the voltage intensity along a number of selected $k$ lines crossing one of two Weyl points, $w_{1}$,
are listed in Fig.\ref{fig:Simu_result}~\textbf{c}-\textbf{h}. We now briefly introduce these figures, while more information can be found in the figure caption. For all of them, the data from simulation and analytic solutions from the model Hamiltonian are plotted by gray dots and dashed red lines, respectively, for comparison. In Fig.\ref{fig:Simu_result}~\textbf{c}, the data from simulation shows that
the surface Weyl states
appear in the gap of the band structure, in good agreement with the results obtained from the model Hamiltonian.
To reveal the local nature of the topological boundary states, the intensity of voltage for the fifth layer to the top layers
is depicted in Fig.\ref{fig:Simu_result}~\textbf{d},
where the surface Weyl states disappear,
because the pulse source on the bottom boundary cannot excite the Weyl surface states located on the top boundary.
Furthermore, according to the model Hamiltonian, the boundary states are contributed only by the node-$c$ and $d$ components, and have vanishing $a$ and $b$ components, which is confirmed by comparing the simulation result in Fig.\ref{fig:Simu_result}~\textbf{f} with that in Fig.\ref{fig:Simu_result}~\textbf{e}.
The intensity of voltage for the bottom four layers is clearly visible only for the component of the $c$ and $d$ nodes plotted in Fig.\ref{fig:Simu_result}~\textbf{f}, while the component of the $a$ and $b$ nodes plotted in Fig.\ref{fig:Simu_result}~\textbf{e} is too weak to be seen.
%
Finally, the band structures of the Weyl states along the $k_{3}$ and $k_{4}$ directions are depicted in Fig.\ref{fig:Simu_result}~\textbf{g} and \textbf{h}, respectively. The linear dispersion relations in the vicinity of the point $w_1$ w.r.t.  all boundary momentum components $\tilde{\boldsymbol{k}}$ show that the point $w_1$ is indeed a Weyl point.

\section*{Conclusions}
In summary, 4D topological states exhibit many interesting phenomena that are markedly different from lower-dimensional topological phases.
Unfortunately, they cannot be realized in condensed-matter materials, which are limited to three spatial dimensions.
In this article we have shown that  periodic electric circuits, composed of  inductors, capacitors, and operational amplifiers, provide
a realistic and ideal platform to create higher-dimensional topological states in the laboratory.
We have explicitly constructed an electric circuit lattice that realizes the 4D spinless topological insulator.
By projecting onto two dimensions, this  4D circuit lattice can readily be implemented on a printed circuit board or an integrated-circuit wafer.
In this way, the higher dimensions of the 4D circuit lattice are faithfully realized through long-ranged lattice connectivity, rather than
by internal degrees of freedom.
Furthermore, the circuit implementation of higher-dimensional topological states has
the advantage of being highly controllable and easily reconfigurable.
This allows, for example, to investigate topological phase transitions, non-Hermitian phenomena, and the effects of nonlinear couplings~\cite{Garcia_LC_PRB2019}.
Using detailed numerical simulations, we have shown that the resonance frequency spectrum of our circuit lattice
exhibits a pair of 3D Weyl boundary states, which is the hallmark of nontrivial topology.

{ Our work opens up the possibility of realizing topological phases in arbitrarily high dimensions, for example
the 5D topological Weyl state~\cite{lian_5D_PRB2016}, or the
6D chiral topological superconductors~\cite{
kitaev_PeriodicTableTopological_2009,
Schnyder-PRB-classification,
ryu_Topological_Table_NJP2010}.
Even topological states on non-orientable surfaces of any dimension could
be realized, such as,
topological phases on M\"obius strips~\cite{Topological_Circuit_PRX_2015}, Klein bottles, or real projective planes.
Other interesting directions for future research concern
the study of quantum effects and
interactions in higher-dimensional topological states.
The former could be simulated by use of periodic Josephson junction arrays~\cite{tsomokos_nori_PRA_10,cosmic_nakamura_PRB_18}.
Moreover, the nonlinearity effect could be achieved by bringing the electronic device into a nonlinear region.
We hope that our work will stimulate further investigations along these lines.}

\section*{Acknowledgements}
We thank Sheng Chang and Dong Zhang for fruitful discussion.
This work was supported by the National Key Research and Development Program of China (No.2017YFA0303402, No.2017YFA0304700),
the National Natural Science Foundation of China (No.11874048, No.11674077), and the GRF (HKU 173057/17P) of Hong Kong.
The numerical calculations in this work have been done on the supercomputing system in the Supercomputing Center of Wuhan University.

\section*{REFERENCES}
\bibliography{refs}

\end{document}